\documentclass[a4paper]{article}

\usepackage{INTERSPEECH2020}

\usepackage{times}
\usepackage{latexsym}

\usepackage{multirow}
\usepackage{graphicx}
\usepackage{float}
\usepackage[caption = false]{subfig}
\usepackage{array}
\usepackage{verbatim}
\usepackage[dvipsnames]{xcolor}
\usepackage{enumitem}

\title{Few-Shot Keyword Spotting With Prototypical Networks}
\name{Archit Parnami, Minwoo Lee}

\address{
  The University of North Carolina at Charlotte}
\email{aparnami@uncc.edu, minwoo.lee@uncc.edu}

\begin{document}

\maketitle
\begin{abstract}
Recognizing a particular command or a keyword, keyword spotting has been widely used in many voice interfaces such as Amazon's Alexa and Google Home. In order to recognize a set of keywords, most of the recent deep learning based approaches use a neural network trained with a large number of samples to identify certain \textit{pre-defined} keywords. This restricts the system from recognizing \textit{new, user-defined} keywords. Therefore, we first formulate this problem as a \textit{few-shot keyword spotting} and approach it using metric learning. To enable this research, we also synthesize and publish a Few-shot Google Speech Commands dataset. We then propose a solution to the few-shot keyword spotting problem using temporal and dilated convolutions on prototypical networks. Our comparative experimental results demonstrate keyword spotting of new keywords using just a small number of samples. 
\end{abstract}

\section{Introduction}
Most smart devices these days have an inbuilt voice recognition system which is mainly used for taking voice input from a user. This requires the voice recognition system to detect specific words (keywords/commands), also known as the Keyword Spotting (KWS) problem. Most approaches use either Large Vocabulary Continuous Speech Recognition (LVCSR) based models \cite{LVCSR1, LVCSR2} or lightweight deep neural network based models \cite{sainath2015convolutional}. The former, LVCSR demands a lot of resource and computation power and hence are deployed in the cloud, raising privacy concerns and latency issues.  The latter models are trained with a set of pre-defined keywords to recognize using thousands of training examples. However, with smart devices becoming more personalized, there is a growing need for such systems 1) to recognize custom or new keywords on-device and 2) to quickly adapt from a small number of user samples as the existing approaches require large number of training samples. Therefore, we attempt to solve this problem of recognizing new keywords given a few samples, hereon referred to as Few-Shot Keyword Spotting (FS-KWS).

Current approaches to KWS involves extracting audio features from the input keyword and then passing it as input to a Deep Neural Network (DNN) for classification \cite{chen2014small, sainath2015convolutional, zhang2017hello, tang2018deep, de2018neural}.  Especially, the use of convolutional neural networks (CNNs) \cite{lecun1998gradient} in adjunction with Mel-frequency Cepstral Coefficients (MFCC) as speech features have shown to produce remarkable results \cite{sainath2015convolutional, chen2014small, de2018neural, choi2019temporal, coucke2019efficient}. 

Due to the data hungry nature of DNNs, recently the field of Few-Shot Learning has gained a lot of attention. Specifically, Few-Shot Classification (FSC) \cite{Chen2019ACL} aims to learn a classifier that can recognize new classes (not seen during training) when given limited, labeled examples for each new class. Broadly there are two approaches to FSC. First, Metric Learning based approaches \cite{koch2015siamese, Vinyals2016MatchingNF,snell2017prototypical} try to learn a good embedding function which can align examples of same class close to each other and far from examples of different class in an embedding space based on a metric (distance function). Second, Optimization based approaches \cite{ravi2016optimization, Finn2017ModelAgnosticMF} attempts to learn good initialization parameters for a classifier such that it can be finetuned using few gradient descent steps on examples from new classes to classify them correctly. Both approaches involve training the classifier with a new set of classes in each training episode such that it will be able to classify another new set of classes at test time.

Previously, \cite{chen2018meta} have attempted to solve FS-KWS using model-agnostic meta learning (MAML) \cite{Finn2017ModelAgnosticMF}, an optimization based approach to FSC. However, since KWS is deployed on small devices with limited computation capability, an optimization based approach that requires fine-tuning may not always be feasible. Hence, we approach FS-KWS using metric learning based approach, specifically using Prototypical Networks \cite{snell2017prototypical} which can perform inference in an end-to-end manner. The following summarizes our main contributions:

\begin{itemize}
    \item We propose a keyword spotting system that can classify new keywords from limited samples by a few-shot formulation of keyword spotting with metric learning.  
    \item We propose a temporally dilated CNN architecture as a better embedding function for FS-KWS.
    \item We release a FS-KWS dataset synthesized from Google\textquotesingle s Speech command dataset \cite{warden2018speech}. To make it more challenging, we also incorporate background noise and detection of silence and unknown (negative) keywords. 
\end{itemize}

\section{Few-Shot Keyword Spotting (FS-KWS) Problem} 

Consider a set $S$ of \textit{user-defined} keywords such that $S =\{(s_i, y_i)\}^{N\times K}_i$ where $s_i$ is a keyword sample (voice input) and $y_i$ is its label. The set $S$ contains $N$ keywords, each keyword having $K$ samples where $K$ is a small number (for ex., 1,2,5). Then given a user query $q$, the objective of FS-KWS system is to classify $q$ into one of $N$ keyword classes. The user-defined keywords in $S$ could be new i.e, never seen before during the training of FS-KWS system. Yet, the system should be able to detect $q$, given $S$.

\begin{figure*}[ht]
\vspace{-0.5cm}
\centering
\includegraphics[keepaspectratio,width=0.9\textwidth]{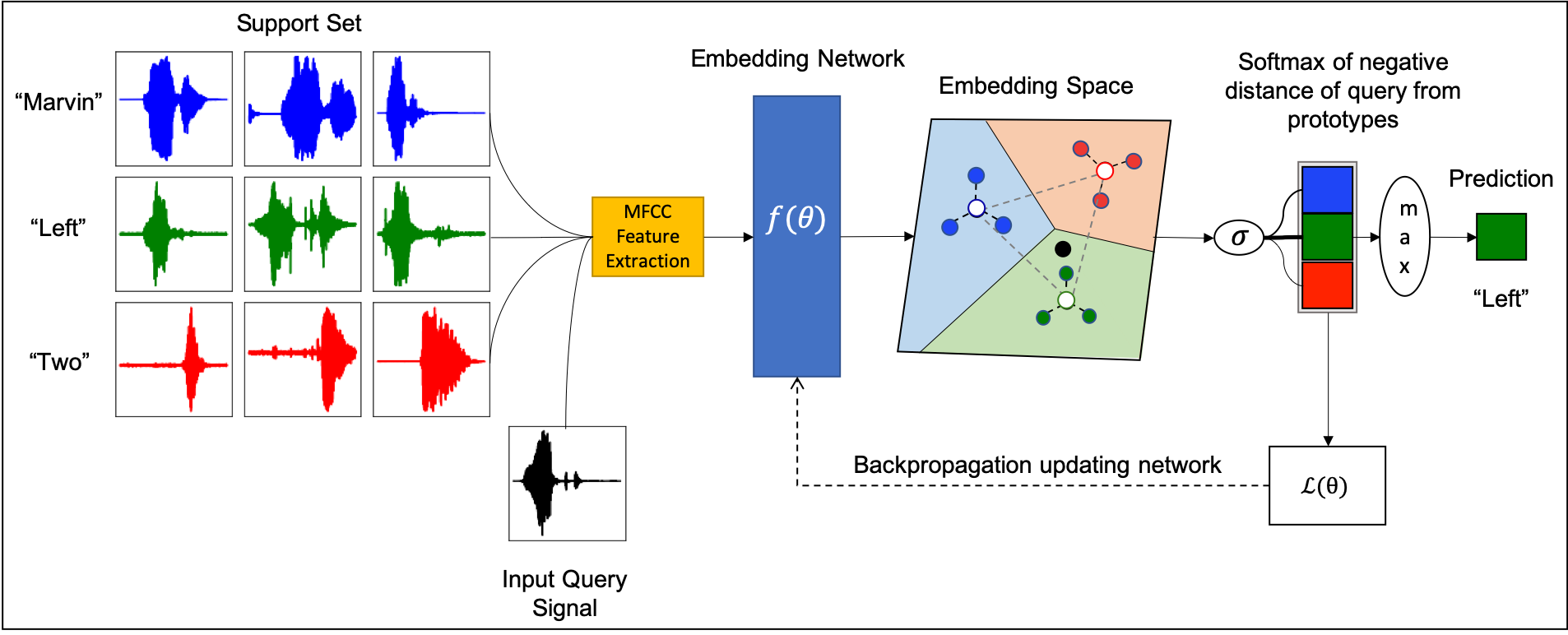}
\caption{Few-Shot Keyword Spotting Pipeline} 
\label{fig:pipeline}
\vspace{-0.5cm}
\end{figure*}

\section{FS-KWS Framework}

\label{sub:framework}
We base our framework (Figure~\ref{fig:pipeline}) on Prototypical Networks \cite{snell2017prototypical} for building the FS-KWS system.
The FS-KWS model is trained on a labeled dataset $D_{train}$ and tested on $D_{test}$. The set of keywords present in $D_{train}$ and $D_{test}$ are disjoint. The test set has only a few labeled samples per keyword. We follow an episodic training paradigm in which each episode the model is trained to solve an $N$-way $K$-Shot FS-KWS task. Each episode $e$ is created by first sampling $N$ categories from the training set and then sampling two sets of examples from these categories: (1) the support set $S_{e} = \{(s_{i},y_{i})\}_{i=1}^{N \times K}$ containing $K$ examples for each of the $N$ categories and (2) the 
query set $Q_{e} = \{(q_{j} , y_{j} )\}_{j =1}^{N \times Q}$ containing $Q$ different examples from the same $N$ categories. The episodic training for FS-KWS  minimizes, for each episode, the loss of the prediction on samples in the query set, given the support set. The model is a parameterized function and the loss is the negative log likelihood of the true class of each query sample:

\begin{equation}\label{eq:loss}
L(\theta) =    -\sum_{t=1}^{\vert Q_{e} \vert} \log P_{\theta} (y_{t} \mid q_{t}, S_{e}),
\end{equation}
where $(q_{t}, y_{t}) \in Q_{e}$ and $S_{e}$ are, respectively, the sampled query and support set at episode $e$ and $\theta$ are the parameters of the model.

Prototypical networks make use of the support set to compute a centroid (prototype) for each category (in the sampled episode) and query samples are classified based on the distance to each prototype. The model is a CNN  $f : \Re^{n_{v}} \to \Re^{n_{p}}$, parameterized by $\theta_{f}$, that learns a $n_{p}$-dimensional space where $n_v$-dimensional input samples of the same category are close and those of different categories are far apart. For every episode $e$, each embedding prototype $p_{c}$ (of category $c$) is computed by averaging the embeddings of all support samples of class $c$:

\begin{equation}
    p_{c} = \frac{1}{\vert S_{e}^c \vert} \sum_{(s_{i}, y_{i}) \in S_{e}^c} f(s_{i}), \nonumber
\end{equation}
where $S_{e}^c \subset S_{e}$ is the subset of support examples belonging to class c. Given a  distance function $d$, the distance of the query $q_{t}$ to each of the class prototypes  $p_{c}$ is calculated. By taking a softmax \cite{bridle1990probabilistic} of the measured (negative) distances, the model produces a distribution over the $N$ categories in each episode:

\vspace{-0.2cm}
 \begin{equation}
    P(y = c \mid q_{t}, S_{e}, \theta)  = \frac{exp(-d(f(q_{t}),p_{c}))}{\sum_{n} exp(-d(f(q_{t}),p_{n}))},\nonumber
\end{equation}
where  metric $d$ is a Euclidean distance and the parameters $\theta$ of the model are updated with stochastic gradient descent by minimizing Equation~(\ref{eq:loss}). Once the training finishes, the parameters $\theta$ of the network are frozen. Then, given any new FS-KWS task, the category corresponding to the maximum $P$ is the predicted category for the input query $q_{t}$.

\begin{figure}[ht]
    \centering
    \subfloat[Input Speech] {\includegraphics[keepaspectratio, width=0.5\linewidth]{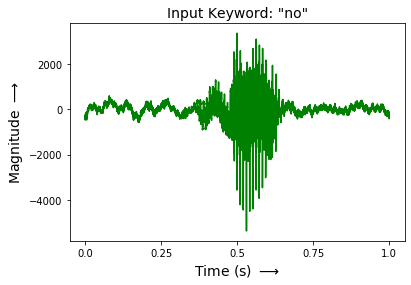}}
    \subfloat[MFCC Features] {\includegraphics[keepaspectratio, width=0.5\linewidth]{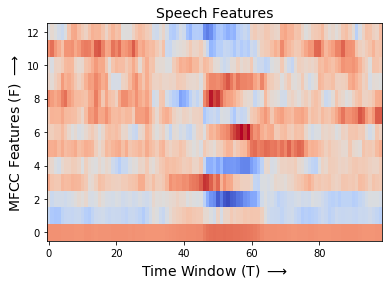}}
    \caption{Example transformation of input speech to MFCC features}
    \label{fig:speech_to_mfcc}
\end{figure}

\subsection{Audio Feature Extraction}

In each episode, we first obtain Mel-frequency Cepstral Coefficients (MFCC) features for all the examples in the support set and the query set which then act as input to the embedding network as shown in Figure \ref{fig:pipeline}. Following \cite{zhang2017hello}, we extract 40 MFCC features  from a speech frame of length 40 \textit{ms} and stride 20 \textit{ms} (see Figure \ref{fig:speech_to_mfcc}).

\begin{figure}[ht]
    \centering
    \includegraphics[keepaspectratio, width=0.9\linewidth]{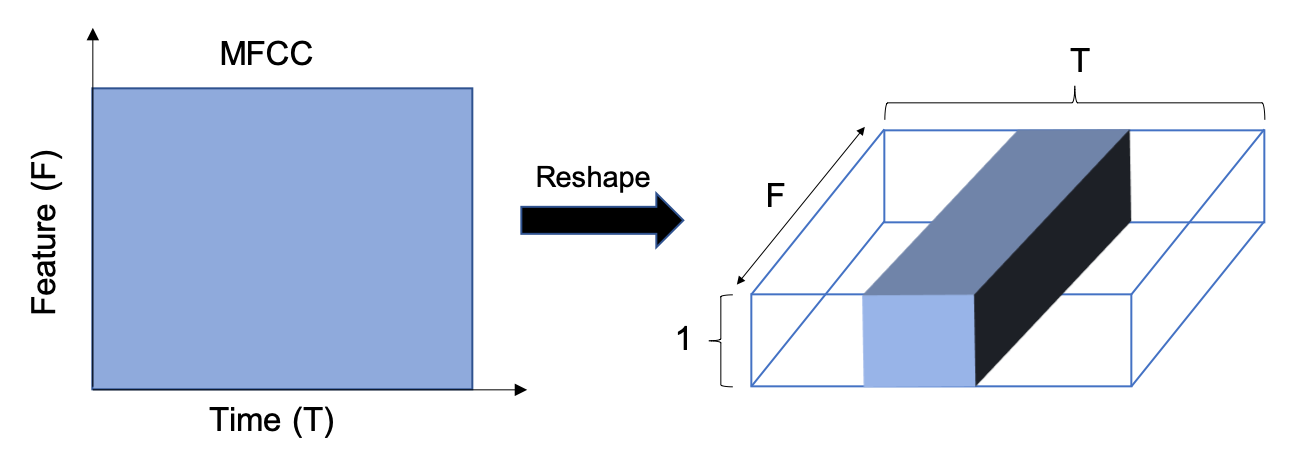}
    \caption{Reshaping MFCC features for time convolution.}
    \label{fig:Reshape}
\end{figure}

\subsection{Embedding Network}
\begin{figure}[ht]
    \centering
    \subfloat[Block] {\includegraphics[keepaspectratio, width=0.4\linewidth]{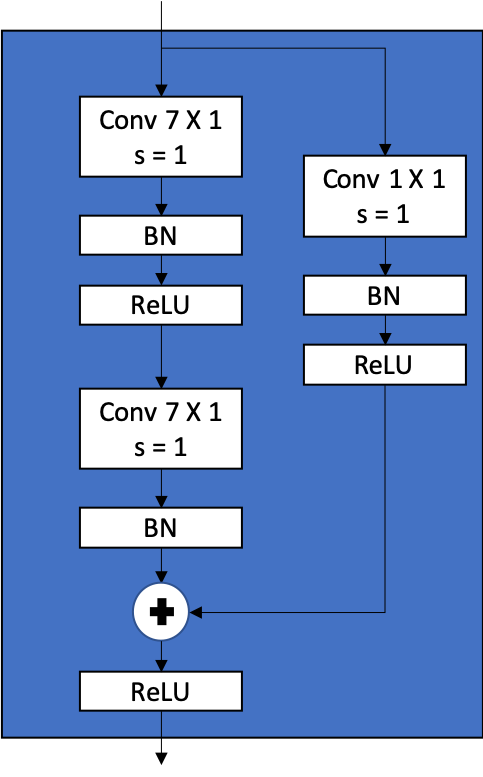}}
    \vspace{0.1\linewidth}
    \subfloat[TD-ResNet7] {\includegraphics[keepaspectratio, width=0.3\linewidth]{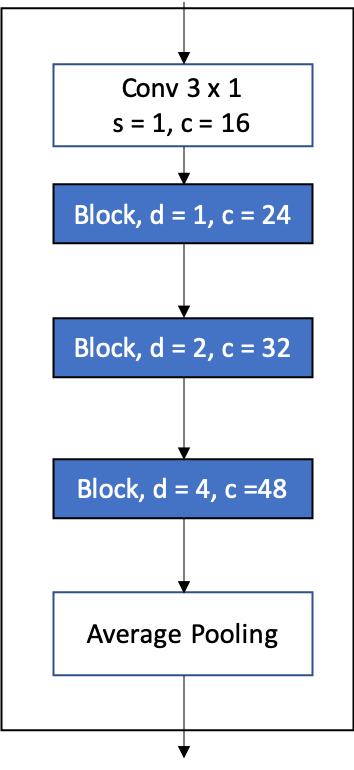}}
    \vspace{-0.8cm}
    \caption{The proposed dilated time convolutional neural network for embedding.}
    \label{fig:TCResNet8Dilated}
    \vspace{-0.5cm}
\end{figure}

Choi et al. \cite{choi2019temporal} demonstrated improved performance on KWS with temporal convolutions by reshaping the input MFCC features (Figure \ref{fig:Reshape}). Also, Cocke et al. \cite{coucke2019efficient} have shown that dilated convolutions are helpful in the processing of keyword signals. Therefore, we combine both the techniques by first reshaping the input MFCC features and then performing temporal convolutions along with dilation. We modify the TC-ResNet8 \cite{choi2019temporal} architecture to reduce the size of the kernel to $7 \times 1$ and use dilation of 1, 2, and 4 with stride 1 in three ResNet blocks respectively. This proposed architecture TD-ResNet7 (Figure \ref{fig:TCResNet8Dilated}) is then used to embed the reshaped input MFCC features (Figure \ref{fig:Reshape}).

\section{Few-Shot Google Speech Command Dataset}

Google's Speech Commands dataset \cite{warden2018speech} has been used previously \cite{zhang2017hello, choi2019temporal} for keyword spotting problem. The dataset has a total of 35 keywords and contains multiple utterances of each keyword by multiple speakers. Each utterance is stored as a one-second (or less) WAVE format file, with the sample data encoded as linear 16-bit single-channel PCM values, at a 16 kHz rate. We curate a FS-KWS dataset from this dataset by performing the following preprocessing steps:

\begin{table}[th!]
\centering
\begin{tabular}{|lllll|}
\hline
\multirow{2}{*}{Keywords} & \multirow{2}{*}{Speakers} & \multicolumn{3}{c|}{Utterances} \\ 
                          &                           & Min      & Max      & Mean     \\ \hline
\multicolumn{5}{|c|}{Core}                                                             \\ \hline
down                      & 1465                      & 1        & 14       & 2.44      \\
zero                      & 1450                      & 1        & 13       & 2.59     \\
seven                     & 1450                      & 1        & 11       & 2.53     \\
nine                      & 1443                      & 1        & 12       & 2.51     \\
five                      & 1442                      & 1        & 19       & 2.58     \\
yes                       & 1422                      & 1        & 20       & 2.6      \\
four                      & 1421                      & 1        & 14       & 2.39     \\
left                      & 1416                      & 1        & 12       & 2.47     \\
stop                      & 1413                      & 1        & 22       & 2.52     \\
six                       & 1411                      & 1        & 14       & 2.55     \\
right                     & 1409                      & 1        & 15       & 2.45     \\
on                        & 1403                      & 1        & 19       & 2.47     \\
three                     & 1401                      & 1        & 11       & 2.43     \\
off                       & 1387                      & 1        & 16       & 2.47     \\
dog                       & 1385                      & 1        & 5        & 1.31     \\
marvin                    & 1378                      & 1        & 6        & 1.33     \\
one                       & 1376                      & 1        & 12       & 2.54     \\
go                        & 1372                      & 1        & 12       & 2.53     \\
no                        & 1368                      & 1        & 18       & 2.59     \\
two                       & 1367                      & 1        & 15       & 2.58     \\
eight                     & 1358                      & 1        & 15       & 2.53     \\
house                     & 1357                      & 1        & 5        & 1.35     \\
wow                       & 1336                      & 1        & 5        & 1.35     \\
happy                     & 1332                      & 1        & 7        & 1.33     \\
bird                      & 1315                      & 1        & 7        & 1.34     \\
cat                       & 1300                      & 1        & 5        & 1.32     \\
up                        & 1291                      & 1        & 17       & 2.53     \\
sheila                    & 1291                      & 1        & 6        & 1.36     \\
bed                       & 1257                      & 1        & 6        & 1.34     \\
tree                      & \textcolor{blue}{1062}                      & 1        & 6        & 1.39     \\ \hline
\multicolumn{5}{|c|}{Unknown}                                                          \\ \hline
visual                    & 412                       & 1        & 7        & 3.57     \\
forward                   & 397                       & 1        & 10       & 3.66     \\
backward                  & 396                       & 1        & 23       & 3.93     \\
follow                    & 387                       & 1        & 11       & 3.76     \\
learn                     & \textcolor{blue}{386}                       & 1        & 24       & 3.69    \\ \hline
\end{tabular}
\caption{Keyword Statistics}
\label{table:keywords}
\vspace{-0.5cm}
\end{table}

\begin{enumerate}
    \item \textbf{Filtering:} We filter out all the utterances which are less than one second. This ensures the consistency of the output MFCC feature matrix obtained from each audio file. 
    \item \textbf{Grouping:} To train our KWS system to detect if an input query is an unknown keyword (not present in $S$), we group our keywords into two categories: \textit{Core} and \textit{Unknown}. Keywords having more than 1000 speakers are considered as \textit{core} words and the rest are put in the category of \textit{unknown} words. 
    \item \textbf{Balancing:} Next, we balance the dataset so that all keywords in a group have the same number of samples. As a result, we have 30 \textit{core} keywords each with 1062 samples and 5 \textit{unknown} keywords each with 386 samples and where all samples for a particular keyword come from a different speaker. 
    \item \textbf{Splitting:} \textit{(a) Core Keywords.} They are randomly split into 20, 5, and 5 sets for training, validation, and testing respectively. Note that here the splits do not have any classes (keywords) in common.
    \textit{(b) Unknown Keywords.} They are used for detecting negative inputs. Since we have only 5 keywords in an unknown category, we utilize them in all three phases of training, validation, and testing. For each keyword in the unknown category, 60\% of its samples are used in training, 20\% for validation, and 20\% for testing. Note that in this case, all the training, validation, and test phases use the same 5 keywords as an unknown class but the samples are still from different speakers.
    \item \textbf{Mixing Background Noise:} The original speech commands dataset \cite{warden2018speech} comes with a collection of sounds (6 WAVE files) that can be mixed with one-second utterances of keywords to simulate background noise. Following \cite{warden2017} implementation  of mixing background noise, small snippets of these files are chosen at random and mixed at a low volume into audio samples during training. The loudness is also chosen randomly, and controlled by a hyper-parameter as a proportion where 0 is silence, and 1 is full volume. In our experiments, we set background volume to 0.1 and conduct experiments with both the presence and absence of background noise.
    \item \textbf{Detecting Silence:} Apart from core classes and unknown class, we curate another class \textit{silence} to detect the absence of keywords. Again following \cite{warden2017} implementation, we randomly sample 1000 one-second long sections of data from background sounds. Since there is never complete silence in real environments, we have to supply examples with quiet and irrelevant audio. We conduct experiments in both the presence and absence of samples from silence class.
    
\end{enumerate}

We provide a script to synthesize this Few-Shot Speech Command dataset at our repository \footnote{https://github.com/ArchitParnami/Few-Shot-KWS}.

\begin{figure*}[ht]
\centering
\includegraphics[keepaspectratio,width=\textwidth]{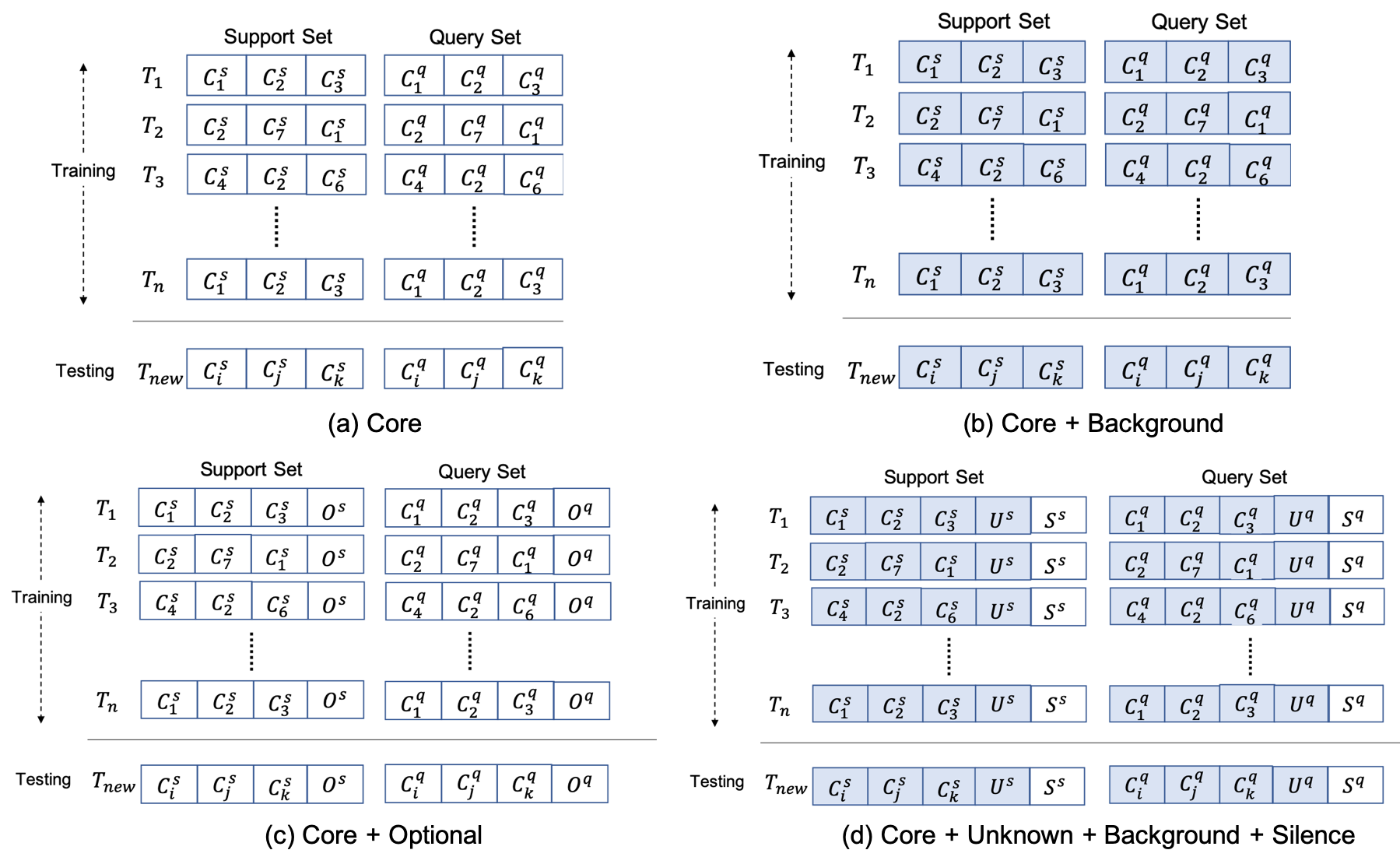}
\caption{Training Cases demonstrated for 3-Way FS-KWS. 
\textbf{(a) Core:} In each task $T_{i}$, 3 \textit{Core} classes are randomly sampled from $D_{train}$. Then for each \textit{Core} class $C_{n}$, \textit{s} support examples $C_{n}^s$ and \textit{q} query examples $C_{n}^q$ are sampled (different from support examples). For testing, a new task $T_{new}$ is constructed which contains new classes $C_{i}, C_{j}, C_{k}$ sampled from $D_{test}$. 
\textbf{(b) Core + Background:} Here each keyword sample is mixed with background noise. 
\textbf{(c) Core + Optional}: An optional class (O) is present along with \textit{Core} classes both during training and testing.  \textbf{(d) Core + Unknown + Background + Silence:} Two optional classes i.e. \textit{Unknown} (U) and \textit{Silence} (S) are present and also the samples are mixed with background noise.
(Note: In our experiments, the position of optional classes in (c) and (d) is random and not always at the last position as presented in this figure)}
\label{fig:cases}
\end{figure*}

\section{Experiments}

\subsection{Training}

\label{sub:training}
To test the effectiveness of our approach, we divide our experiments in four cases (Figure \ref{fig:cases}): 

\begin{enumerate}[label=(\alph*)]
    \item \textbf{Core} - Pure Keyword Detection: Both during training and testing, the keyword samples in the support ($S$) and query ($Q$) sets  are from \textit{core keywords} and without any background noise.
    
    \item  \textbf{Core + Background}: Same as (a), except the keyword samples are now mixed with random background noise.
    
    \item \textbf{Core + Optional}: To account for scenarios when the input query is not from any of the keywords present in the provided support set or when there is simply no input, we train and test in presence of an optional class. This optional class is \textit{unknown keywords} when we want to detect negatives and is \textit{silence} when we want to detect the absence of any spoken keywords. 
    
    \item \textbf{Core + Unknown + Silence + Background}: Samples from both the optional classes i.e, \textit{Unknown} and \textit{Silence} are present and are also mixed with background noise. This case simulates more realistic scenarios when input is often mixed with background noise and could be an unknown word or just silence. 
    
\end{enumerate}

In each of the above cases, we train and test in a $N$-way $K$-shot manner where $N$ refers to the number of \textit{core} classes and $K$ refers to the number of training examples per class in each episode as explained in Section \ref{sub:framework}. In cases where an optional class (\textit{Silence} or \textit{Unknown}) is used, we add $K$ support examples for the optional class in the support sets both during training and testing.  We perform episodic training as suggested in \cite{snell2017prototypical} and train all our models for 200 epochs where each epoch has 200 training episodes and 100 validation (test) episodes. We use SGD with Adam \cite{kingma2014adam} and an initial learning rate of $10^{-3}$ and cut the learning rate in half every 20 epochs. We conduct experiments with  $N = \{2,4\}$ and $K=\{1,5\}$ for all the mentioned cases. The model is trained on the loss computed from 5 queries per class in each episode and evaluated more strictly with 15 queries per class during testing.

\begin{figure*} [ht]
    \centering
    \subfloat[Core]{\includegraphics[keepaspectratio,width=0.5\linewidth]{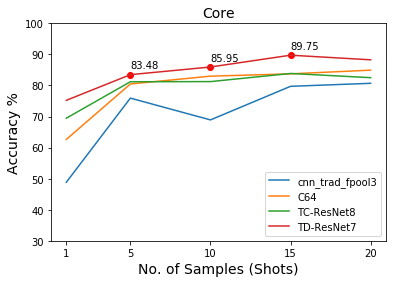}}
    \subfloat[Core + Background]{\includegraphics[keepaspectratio,width=0.5\linewidth]{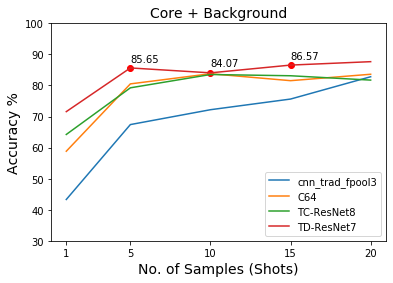}}
    \\
    \subfloat[Core + Unknown]{\includegraphics[keepaspectratio,width=0.5\linewidth]{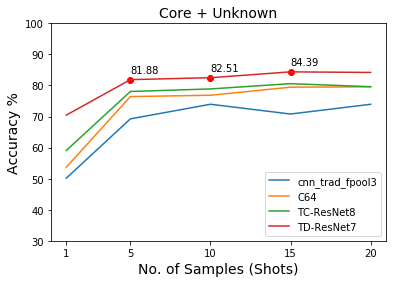}}
    \subfloat[Core + Unknown + Background + Silence]{\includegraphics[keepaspectratio,width=0.5\linewidth]{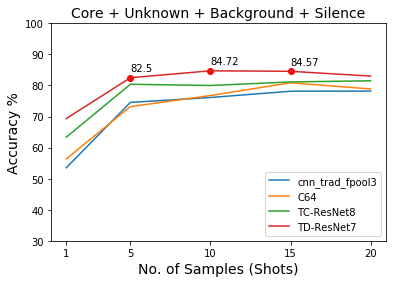}}
    \caption{Comparing test accuracy of embedding network architectures on 4-way FS-KWS as we increase the number of support examples. The results are presented for all the four cases mentioned in section \ref{sub:training}}
    \label{fig:performance}
\end{figure*}{}

\subsection{Baselines}

As we formulate and propose a new FS-KWS problem, there is a lack of prior research and standard FS-KWS dataset. Thus, to show the effectiveness of the proposed framework, we employ three different existing architectures as embedding network in our FS-KWS framework to examine the performance of the proposed approach. Following are the baseline embedding networks:

\begin{itemize}
    \item \textbf{cnn\_trad\_fpool3} \cite{sainath2015convolutional} was originally proposed for KWS problem. It has two convolutional layers followed by a linear, a dense, and a softmax layer. We use the output of the dense layer as network embeddings.  
    \item \textbf{C64} \cite{snell2017prototypical} is the original 4-layer CNN used in Prototypical Networks for doing few-shot image classification on miniImageNet \cite{Vinyals2016MatchingNF}.
    \item \textbf{TC-ResNet8} \cite{choi2019temporal} has demonstrated great results on KWS. We remove the last fully connected and softmax layer and use the remaining architecture as our embedding network in FS-KWS framework. 
\end{itemize}

\subsection{Results}
Table \ref{table:results} lists the results for the three baselines and our proposed architecture on experiments mentioned in Section~\ref{sub:training}. Given a new 2-way-5-shot KWS task with keywords \textbf{not seen} during the training, our TD-ResNet7 model can classify an input query with $\sim$94\% accuracy with the proposed FW-KWS pipeline. 
This is not even feasible with classical deep learning solutions withou FS-KWS formulation. 

The TD-ResNet7 architecture also outperforms all the existing baselines architectures on all the test cases except in \textit{(b) Core + Background} where the performance of TC-ResNet8 on 2-way 5-shot KWS is slightly better but the difference is not significant ($p=0.36$ while ANOVA for others presents $p\ll0.05$). These results are illustrated in Figure \ref{fig:performance}. As we increase the number of shots (samples per class), the overall performance improves for all architectures, yet the TD-ResNet7 architecture consistently outperforms other baselines. All the accuracy results are averaged over 100 test episodes and are reported with 95\% confidence intervals.

\begin{table*}[th!]
\centering
\begin{tabular}{|c|c|c|c|c|c|}
\hline
\multirow{2}{*}{\textbf{Case}}                                                                         & \multirow{2}{*}{\textbf{\begin{tabular}[c]{@{}c@{}}Embedding\\ Network\end{tabular}}} & \multicolumn{2}{c|}{\textbf{2-way Acc.}}              & \multicolumn{2}{c|}{\textbf{4-way Acc.}}              \\ \cline{3-6} 
                                                                                                       &                                                                                       & 1-shot                    & 5-shot                    & 1-shot                    & 5-shot                    \\ \hline
\multirow{4}{*}{core}                                                                                  & cnn\_trad\_fpool3                                                                     & 69.23 $\pm$ 0.03          & 87.07 $\pm$ 0.02          & 48.83 $\pm$ 0.02          & 75.93 $\pm$ 0.01          \\
                                                                                                       & C64                                                                                   & 77.20 $\pm$ 0.03          & 89.97 $\pm$ 0.02          & 62.63 $\pm$ 0.02          & 80.48 $\pm$ 0.01          \\
                                                                                                       & TC-ResNet8                                                                             & 82.70 $\pm$ 0.03          & 89.00 $\pm$ 0.02          & 69.47 $\pm$ 0.02          & 81.20 $\pm$ 0.01          \\
                                                                                                       & TD-ResNet7 (ours)                                                                      & \textbf{85.43 $\pm$ 0.03} & \textbf{94.10 $\pm$ 0.01} & \textbf{75.22 $\pm$ 0.02} & \textbf{83.48 $\pm$ 0.02} \\ \hline
\multirow{4}{*}{\begin{tabular}[c]{@{}c@{}}core \\ + \\ background\end{tabular}}                       & cnn\_trad\_fpool3                                                                     & 69.53 $\pm$ 0.04          & 86.8 $\pm$ 0.02           & 43.3 $\pm$ 0.02           & 67.42 $\pm$ 0.01          \\
                                                                                                       & C64                                                                                   & 78.30 $\pm$ 0.03          & 90.03 $\pm$ 0.02          & 58.83 $\pm$ 0.02          & 80.52 $\pm$ 0.01          \\
                                                                                                       & TC-ResNet8                                                                             & 77.40 $\pm$ 0.03          & \textbf{91.40 $\pm$ 0.02} & 64.23 $\pm$ 0.02          & 79.25 $\pm$ 0.01          \\
                                                                                                       & TD-ResNet7 (ours)                                                                      & \textbf{82.23 $\pm$ 0.03} & 91.00 $\pm$ 0.02          & \textbf{71.58 $\pm$ 0.02} & \textbf{85.65 $\pm$ 0.01} \\ \hline
\multirow{4}{*}{\begin{tabular}[c]{@{}c@{}}core \\ + \\ unknown\end{tabular}}                          & cnn\_trad\_fpool3                                                                     & 58.33 $\pm$ 0.03          & 78.36 $\pm$ 0.02          & 50.15 $\pm$ 0.02          & 69.25 $\pm$ 0.01          \\
                                                                                                       & C64                                                                                   & 63.42 $\pm$ 0.03          & 78.47 $\pm$ 0.02          & 53.69 $\pm$ 0.02          & 76.43 $\pm$ 0.01          \\
                                                                                                       & TC-ResNet8                                                                             & 68.84 $\pm$ 0.03          & 80.49 $\pm$ 0.02          & 59.08 $\pm$ 0.02          & 78.07 $\pm$ 0.01          \\
                                                                                                       & TD-ResNet7 (ours)                                                                      & \textbf{77.24 $\pm$ 0.02} & \textbf{87.22 $\pm$ 0.01} & \textbf{70.45 $\pm$ 0.02} & \textbf{81.88 $\pm$ 0.01} \\ \hline
\multirow{4}{*}{\begin{tabular}[c]{@{}c@{}}core +  \\ unknown +\\ background +\\ silence\end{tabular}} & cnn\_trad\_fpool3                                                                     & 67.43 $\pm$ 0.02          & 82.32 $\pm$ 0.01          & 53.51 $\pm$ 0.02          & 74.54 $\pm$ 0.01          \\
                                                                                                       & C64                                                                                   & 65.83 $\pm$ 0.02          & 81.15 $\pm$ 0.01          & 56.38 $\pm$ 0.01          & 73.20 $\pm$ 0.01          \\
                                                                                                       & TC-ResNet8                                                                             & 78.63 $\pm$ 0.02          & 85.98 $\pm$ 0.01          & 63.37 $\pm$ 0.02          & 80.39 $\pm$ 0.01          \\
                                                                                                       & TD-ResNet7 (ours)                                                                      & \textbf{82.77 $\pm$ 0.02} & \textbf{89.45 $\pm$ 0.01} & \textbf{69.34 $\pm$ 0.01} & \textbf{82.50 $\pm$ 0.01} \\ \hline
\end{tabular}

\caption{Performance comparison of different embedding networks when plugged into FS-KWS pipeline for 4 different cases.}
\label{table:results}
\end{table*}

\section{Conclusion}
In this work, we attempted to solve the keyword spotting problem using only limited samples from each keyword. We demonstrated that using prototypical networks with our proposed embedding model which uses temporal and dilated convolutions, can produce significant results with only few examples. We also synthesis and release a Few-Shot Google Speech command dataset for future research on Few-Shot Keyword Spotting.

\bibliographystyle{IEEEtran}

\bibliography{main}

\begin{thebibliography}{10}
\providecommand{\url}[1]{#1}
\csname url@samestyle\endcsname
\providecommand{\newblock}{\relax}
\providecommand{\bibinfo}[2]{#2}
\providecommand{\BIBentrySTDinterwordspacing}{\spaceskip=0pt\relax}
\providecommand{\BIBentryALTinterwordstretchfactor}{4}
\providecommand{\BIBentryALTinterwordspacing}{\spaceskip=\fontdimen2\font plus
\BIBentryALTinterwordstretchfactor\fontdimen3\font minus
  \fontdimen4\font\relax}
\providecommand{\BIBforeignlanguage}[2]{{%
\expandafter\ifx\csname l@#1\endcsname\relax
\typeout{** WARNING: IEEEtran.bst: No hyphenation pattern has been}%
\typeout{** loaded for the language `#1'. Using the pattern for}%
\typeout{** the default language instead.}%
\else
\language=\csname l@#1\endcsname
\fi
#2}}
\providecommand{\BIBdecl}{\relax}
\BIBdecl

\bibitem{LVCSR1}
P.~{Motlicek}, F.~{Valente}, and I.~{Szoke}, ``Improving acoustic based keyword
  spotting using lvcsr lattices,'' in \emph{2012 IEEE International Conference
  on Acoustics, Speech and Signal Processing (ICASSP)}, 2012, pp. 4413--4416.

\bibitem{LVCSR2}
D.~{Can} and M.~{Saraclar}, ``Lattice indexing for spoken term detection,''
  \emph{IEEE Transactions on Audio, Speech, and Language Processing}, vol.~19,
  no.~8, pp. 2338--2347, 2011.

\bibitem{sainath2015convolutional}
T.~N. Sainath and C.~Parada, ``Convolutional neural networks for
  small-footprint keyword spotting,'' in \emph{Sixteenth Annual Conference of
  the International Speech Communication Association}, 2015.

\bibitem{chen2014small}
G.~Chen, C.~Parada, and G.~Heigold, ``Small-footprint keyword spotting using
  deep neural networks,'' in \emph{2014 IEEE International Conference on
  Acoustics, Speech and Signal Processing (ICASSP)}.\hskip 1em plus 0.5em minus
  0.4em\relax IEEE, 2014, pp. 4087--4091.

\bibitem{zhang2017hello}
Y.~Zhang, N.~Suda, L.~Lai, and V.~Chandra, ``Hello edge: Keyword spotting on
  microcontrollers,'' \emph{arXiv preprint arXiv:1711.07128}, 2017.

\bibitem{tang2018deep}
R.~Tang and J.~Lin, ``Deep residual learning for small-footprint keyword
  spotting,'' in \emph{2018 IEEE International Conference on Acoustics, Speech
  and Signal Processing (ICASSP)}.\hskip 1em plus 0.5em minus 0.4em\relax IEEE,
  2018, pp. 5484--5488.

\bibitem{de2018neural}
D.~C. de~Andrade, S.~Leo, M.~L. D.~S. Viana, and C.~Bernkopf, ``A neural
  attention model for speech command recognition,'' \emph{arXiv preprint
  arXiv:1808.08929}, 2018.

\bibitem{lecun1998gradient}
Y.~LeCun, L.~Bottou, Y.~Bengio, P.~Haffner \emph{et~al.}, ``Gradient-based
  learning applied to document recognition,'' \emph{Proceedings of the IEEE},
  vol.~86, no.~11, pp. 2278--2324, 1998.

\bibitem{choi2019temporal}
S.~Choi, S.~Seo, B.~Shin, H.~Byun, M.~Kersner, B.~Kim, D.~Kim, and S.~Ha,
  ``Temporal convolution for real-time keyword spotting on mobile devices,''
  \emph{arXiv preprint arXiv:1904.03814}, 2019.

\bibitem{coucke2019efficient}
A.~Coucke, M.~Chlieh, T.~Gisselbrecht, D.~Leroy, M.~Poumeyrol, and T.~Lavril,
  ``Efficient keyword spotting using dilated convolutions and gating,'' in
  \emph{ICASSP 2019-2019 IEEE International Conference on Acoustics, Speech and
  Signal Processing (ICASSP)}.\hskip 1em plus 0.5em minus 0.4em\relax IEEE,
  2019, pp. 6351--6355.

\bibitem{Chen2019ACL}
W.-Y. Chen, Y.-C. Liu, Z.~Kira, Y.-C.~F. Wang, and J.-B. Huang, ``A closer look
  at few-shot classification,'' \emph{ArXiv}, vol. abs/1904.04232, 2019.

\bibitem{koch2015siamese}
G.~Koch, R.~Zemel, and R.~Salakhutdinov, ``Siamese neural networks for one-shot
  image recognition,'' in \emph{ICML deep learning workshop}, vol.~2, 2015.

\bibitem{Vinyals2016MatchingNF}
O.~Vinyals, C.~Blundell, T.~P. Lillicrap, K.~Kavukcuoglu, and D.~Wierstra,
  ``Matching networks for one shot learning,'' in \emph{NIPS}, 2016.

\bibitem{snell2017prototypical}
J.~Snell, K.~Swersky, and R.~Zemel, ``Prototypical networks for few-shot
  learning,'' in \emph{Advances in Neural Information Processing Systems},
  2017, pp. 4077--4087.

\bibitem{ravi2016optimization}
S.~Ravi and H.~Larochelle, ``Optimization as a model for few-shot learning,''
  in \emph{ICLR}, 2017.

\bibitem{Finn2017ModelAgnosticMF}
C.~Finn, P.~Abbeel, and S.~Levine, ``Model-agnostic meta-learning for fast
  adaptation of deep networks,'' in \emph{ICML}, 2017.

\bibitem{chen2018meta}
Y.~Chen, T.~Ko, L.~Shang, X.~Chen, X.~Jiang, and Q.~Li, ``Meta learning for
  few-shot keyword spotting,'' \emph{arXiv preprint arXiv:1812.10233}, 2018.

\bibitem{warden2018speech}
P.~Warden, ``Speech commands: A dataset for limited-vocabulary speech
  recognition,'' \emph{arXiv preprint arXiv:1804.03209}, 2018.

\bibitem{bridle1990probabilistic}
J.~S. Bridle, ``Probabilistic interpretation of feedforward classification
  network outputs, with relationships to statistical pattern recognition,'' in
  \emph{Neurocomputing}.\hskip 1em plus 0.5em minus 0.4em\relax Springer, 1990,
  pp. 227--236.

\bibitem{warden2017}
\BIBentryALTinterwordspacing
P.~Warden, \emph{Launching the speech commands dataset.}, 2017. [Online].
  Available:
  \url{https://ai.googleblog.com/2017/08/launching-speech-commands-dataset.html}
\BIBentrySTDinterwordspacing

\bibitem{kingma2014adam}
D.~P. Kingma and J.~Ba, ``Adam: A method for stochastic optimization,''
  \emph{arXiv preprint arXiv:1412.6980}, 2014.

\end{thebibliography}

\end{document}